\documentclass[sigconf,natbib=true]{acmart}
\usepackage{CJKutf8}
\usepackage{float}
\usepackage{multirow}
\AtBeginDocument{%
  \providecommand\BibTeX{{%
    \normalfont B\kern-0.5em{\scshape i\kern-0.25em b}\kern-0.8em\TeX}}}

\copyrightyear{2022}
\acmYear{2022}
\setcopyright{acmlicensed}\acmConference[SIGIR '22]{Proceedings of the 45th International ACM SIGIR Conference on Research and Development in Information Retrieval}{July 11--15, 2022}{Madrid, Spain}
\acmBooktitle{Proceedings of the 45th International ACM SIGIR Conference on Research and Development in Information Retrieval (SIGIR '22), July 11--15, 2022, Madrid, Spain}
\acmPrice{15.00}
\acmDOI{10.1145/3477495.3531736}
\acmISBN{978-1-4503-8732-3/22/07}

\begin{document}
\begin{CJK}{UTF8}{gbsn}

\title{Multi-CPR: A Multi Domain Chinese Dataset for Passage Retrieval}

\author{Dingkun Long, Qiong Gao, Kuan Zou, Guangwei Xu, Pengjun Xie, Ruijie Guo, \\ Jian Xu, Guanjun Jiang, Luxi Xing, Ping Yang}
\email{dingkun.ldk, gaoqiong.gao, zoukuan.zk, kunka.xgw, chengchen.xpj, jian.xujian@alibaba-inc.com}
\email{ruijie.guo, luxi.xlx, yangping.yangping, guanj.jianggj@alibaba-inc.com}
\affiliation{%
  \institution{Alibaba Group}
  \city{Hangzhou}
  \country{China}
}
\renewcommand{\shortauthors}{Dingkun Long and Qiong Gao, et al.}
\begin{abstract}
Passage retrieval is a fundamental task in information retrieval (IR) research, which has drawn much attention recently. In the English field, the availability of large-scale annotated dataset (e.g, MS MARCO) and the emergence of deep pre-trained language models (e.g, BERT) has resulted in a substantial improvement of existing passage retrieval systems. However, in the Chinese field, especially for specific domains, passage retrieval systems are still immature due to quality-annotated dataset being limited by scale. Therefore, in this paper, we present a novel multi-domain Chinese dataset for passage retrieval (Multi-CPR). The dataset is collected from three different domains, including E-commerce, Entertainment video and Medical. Each dataset contains millions of passages and a certain amount of human annotated query-passage related pairs. We implement various representative passage retrieval methods as baselines. We find that the performance of retrieval models trained on dataset from general domain will inevitably decrease on specific domain. Nevertheless, a passage retrieval system built on in-domain annotated dataset can achieve significant improvement, which indeed demonstrates the necessity of domain labeled data for further optimization. We hope the release of the Multi-CPR dataset could benchmark Chinese passage retrieval task in specific domain and also make advances for future studies. 
\end{abstract}

\maketitle


\section{Introduction}
Large scale passage retrieval is an important problem in information retrieval research field. Passage retrieval is often regarded as a prerequisite to downstream tasks and applications like open-domain question answering~\cite{karpukhin2020dense,li2021encoder}, machine reading comprehension~\cite{rajpurkar2016squad,nishida2018retrieve} and web search systems~\cite{cai2004block}, etc. Recent advances in deep learning have allowed state of the art performance on passage retrieval task compared to conventional statistical models~\cite{karpukhin2020dense,gao2021coil,qu2021rocketqa,gao2021condenser,gao2021unsupervised}. However, these deep neural models usually contain millions of parameters that necessitate a large amount of training data. As such, high-quality public available benchmark dataset is critical for research progress with a deep-model fashion for the passage retrieval task.

In the English field, we observed that large, high-quality dataset enables the community rapidly develop new models for passage retrieval task, and at the same time, the research on model architecture also obtains a more deep understanding. As mentioned above, passage retrieval mainly serves downstream tasks such as question answering and machine reading comprehension. Therefore, existing datasets are also constructed based on the above two tasks. In term of question answering, there are several benchmark datasets like TREC QA~\cite{wang2007jeopardy}, WikiPassageQA~\cite{Cohen2018WikiPassageQAAB} and InsuranceQA~\cite{Feng2015ApplyingDL}. For machine reading comprehension task, representative datasets including SQuAD~\cite{rajpurkar-etal-2016-squad}, MS MARCO~\cite{Campos2016MSMA}, CNN /Daily News~\cite{Hermann2015TeachingMT} provide good benchmarks. In summary, dataset in the English field is relatively mature in terms of data scale and domain richness. On the other hand, in the field of Chinese, although some information retrieval datasets and machine reading comprehension datasets have been released in recent years like Sogou-QCL~\cite{Zheng2018sougou}, Dureader~\cite{he-etal-2018-dureader} and SC-MRC~\cite{cui-etal-2020-sentence}, these datasets are mainly concentrated in the general domain, and dataset that can be adopted for specific domain passage retrieval research is still in shortage.

To push forward the quality and variety of Chinese passage retrieval dataset, we present Multi-CPR. There are three main properties of Multi-CPR: a) Multi-CPR is the first dataset that covers multiple specific domains for Chinese passage retrieval, including E-commerce, Entertainment video and Medical. There is a high degree of differentiation within the three domains. Furthermore, Only one (Medical) of these domains has been studied in previous research~\cite{zhang2020conceptualized}. b) Multi-CPR is the largest domain specific Chinese passage retrieval dataset. For each domain, Multi-CPR contains millions of passages (e,g. 1,002,822 passages for the E-commerce domain) and sufficient human annotated query-passage related pairs. More detailed statistics of Multi-CPR and annotated examples can be found in Table \ref{tab:daa-statistics} and  Section \ref{section:data-annotation}. c) All Queries and passages in Multi-CPR are collected from real search engine systems within Alibaba Group. The authenticity of the samples allows Multi-CPR to meet the needs of both academia and industry fields.

As an attempt to tackle Multi-CPR and provide strong baselines, we implement various representative passage retrieval methods including both sparse and dense models. For the sparse models, except for the basic BM25 method, we also verified that previously proposed optimization methods based on the sparse strategy can indeed achieve significant improvement compared to the BM25 baseline (e,g, doc2query method). For the dense models, we mainly implemented methods based on the DPR model. Similarly, we also made some optimizations based on the DPR model. Compared to the sparse models, we found that the retrieval performance of the dense models trained on labeled dataset can be greatly improved. This observation empirically confirms the value of annotated data. In further, we verified that the retrieval-then-reranking two-stage framework based on the BERT model can further improve the overall retrieval performance on all three datasets in Multi-CPR, which once again corroborates the quality of Multi-CPR.

In summary, the major contributions of this paper are threefold:
\begin{itemize}
    \item We present Multi-CPR, the largest-scale Chinese multi domain passage retrieval dataset collected from practical search engine systems, and it covers E-commence, Entaitement vedio and Medical domain.
    \item We conduct an in-depth analysis on Multi-CPR. Based on Multi-CPR, we have analyzed the characteristics of different passage retrieval methods along with their optimization strategies associated, which enables us to have a deeper understanding of Chinese passage retrieval task in specific domain.
    \item We implement various representative methods as baselines and show the performance of existing methods on Multi-CPR, which provides an outlook for future research.
\end{itemize}

\section{Related Work}
\label{sec:related-work}
\noindent{\textbf{Passage Retrieval}} Passage retrieval task aims to recall all potentially relevant passages from a large corpus given an information-seeking query. In practical, passage retrieval is often an important step in other information retrieval tasks~\cite{cai2004block}. 
Traditional passage retrieval systems usually rely on term-based retrieval models like BM25~\cite{robertson2009probabilistic}. Recently, with the rapid development in text representation learning research~\cite{bengio2013representation} and deep pre-trained language models~\cite{kenton2019bert,liu2019roberta,yang2019xlnet,he2020deberta}, dense retrieval combined with pre-trained language models, has become a popular paradigm to improve retrieval performance~\cite{karpukhin2020dense,gao2021unsupervised,qu2021rocketqa}. In general, dense models significantly outperform traditional term-based retrieval models in terms of effectiveness and benefit downstream tasks.

In a basic concept, the core problem of passage retrieval is how to form the text representation and then compute text similarity. Thus, based on the text representation type and corpus index mode, passage retrieval models can be roughly categorized into two main classes. Sparse retrieval Models: improving retrieval by obtaining semantic-captured sparse representations and indexing them with the inverted index for efficient retrieval; Dense Retrieval Models: converting query and passage into continuous embedding representations and turning to approximate nearest neighbor (ANN) algorithms for fast retrieval~\cite{fan2021pre}.

\begin{table}[t]
\centering
\renewcommand{\arraystretch}{1.0}
\caption{Example of annotated query-passage related pairs in three different domains.}
\label{table:example-three-domain}
\begin{tabular}{c|c|p{4.5cm}}
\toprule
\multirow{2}{*}{E-commerce} & Query   & 尼康z62 (\color[HTML]{3166FF}{Nikon z62}) \\ \cmidrule(l){2-3} 
                            & Passage & Nikon/尼康二代全画幅微单机身Z62 Z72 24-70mm套机 (\color[HTML]{3166FF}{Nikon/Nikon II, full-frame micro-single camera, body Z62 Z72 24-70mm set})\\ \midrule
\multirow{2}{*}{\begin{tabular}[c]{@{}c@{}} Entertainment\\ video\end{tabular}} & Query   & 海神妈祖 (\color[HTML]{3166FF}{Ma-tsu, Goddess of the Sea}) \\ \cmidrule(l){2-3} 
                            & Passage & 海上女神妈祖 (\color[HTML]{3166FF}{Ma-tsu, Goddess of the Sea} ) \\ \midrule
\multirow{2}{*}{Medical} & Query   & 大人能把手放在睡觉婴儿胸口吗 (\color[HTML]{3166FF}{Can adults put their hands on the chest of a sleeping baby?})\\ \cmidrule(l){2-3} 
                            & Passage & 大人不能把手放在睡觉婴儿胸口，对孩子呼吸不好，要注意 (\color[HTML]{3166FF}{Adults should not put their hands on the chest of a sleeping baby as this is not good for the baby's breathing.}) \\ \bottomrule
\end{tabular}
\end{table}

For the above two types of models, the current optimization directions are not the same. Specifically, Sparse retrieval models focus on improving retrieval performance by either enhancing the bag-of-words (BoW) representations in classical term-based methods or mapping input texts into latent space (e,g. doc2query~\cite{Nogueira2019FromDT}, query expansion~\cite{cui2002probabilistic} and document expansion~\cite{nogueira2019document}). The sparse representation has attracted great attention as it can be easily integrated into the inverted index for efficient retrieval. Recently, With the development of deep neural networks, pre-trained language models have been widely employed to improve the capacity of sparse retrieval models, including term re-weighting~\cite{dai2019context,dai2019deeper}, sparse representation learning~\cite{jang2021uhd,yamada2021efficient}, etc. The mainstream of existing studies on improving the performance of dense retrieval models can be roughly divided into three groups. 1) Designing more powerful pre-trained language model architectures for the passage retrieval task and then improving the quality of sentence representation. For example, ~\cite{gao2021condenser,gao2021unsupervised} proposed the Condenser family models. 2) Applying pre-training methods for dense retrieval is to use pre-trained models as encoders, and then fine-tuned with labeled dataset. In the fine-tuning stage, existing research work attempted to select more reasonable hard negative samples~\cite{xiong2020approximate,zhan2021optimizing}. 3) Existing state-of-the-art retrieval systems usually leverage a two-stage (retrieval-then-reranking) framework. Different from the previous traditional pipeline model, ~\cite{ren2021rocketqav2} and ~\cite{zhang2022adversarial} proposed to better leverage the feedback from reranker to promote the performance of retrieval stage via joint learning and adversarial learning respectively. 

\noindent{\textbf{Related Datasets}} As mentioned above, the emergence of large-scale high-quality labeled data has greatly promoted the optimization process of passage retrieval models. Among all these datasets, MS MARCO~\cite{Campos2016MSMA} is the most representative dataset in the English field. MS MARCO is a passage and document ranking dataset introduced by Microsoft. The passage ranking task focuses on ranking passages from a collection of about 8.8 million, which are gathered from Bing’s results to real-world queries. About $808$ thousand queries paired with relevant passages are provided for supervised training. Each query is associated with sparse relevance judgments of one (or very few) passages marked as relevant and no passages explicitly indicated as irrelevant. 

In the Chinese field, there are some datasets built based on web page retrieval systems, for example, Sougou-QCL~\cite{Zheng2018sougou}. The Sogou-QCL dataset was created to support research on information retrieval and related human language technologies. The dataset consists of 537,366 queries, more than 9 million Chinese web pages, and five kinds of relevance labels assessed by click models. However, this dataset is concentrated in the general domain, and the labels are obtained based on click behavior, rather than human annotation. Dureader is a recently released large-scale MRC dataset in Chinese~\cite{he-etal-2018-dureader}. The data distribution is mainly concentrated in the general domain. It can be converted into an information retrieval dataset. Although there are some general domain dataset available, Chinese passage retrieval annotated dataset in specific domain is still in shortage.

\begin{figure*}[t]
  \centering
  \label{fig:annotation-plat}
  \includegraphics[width=\linewidth]{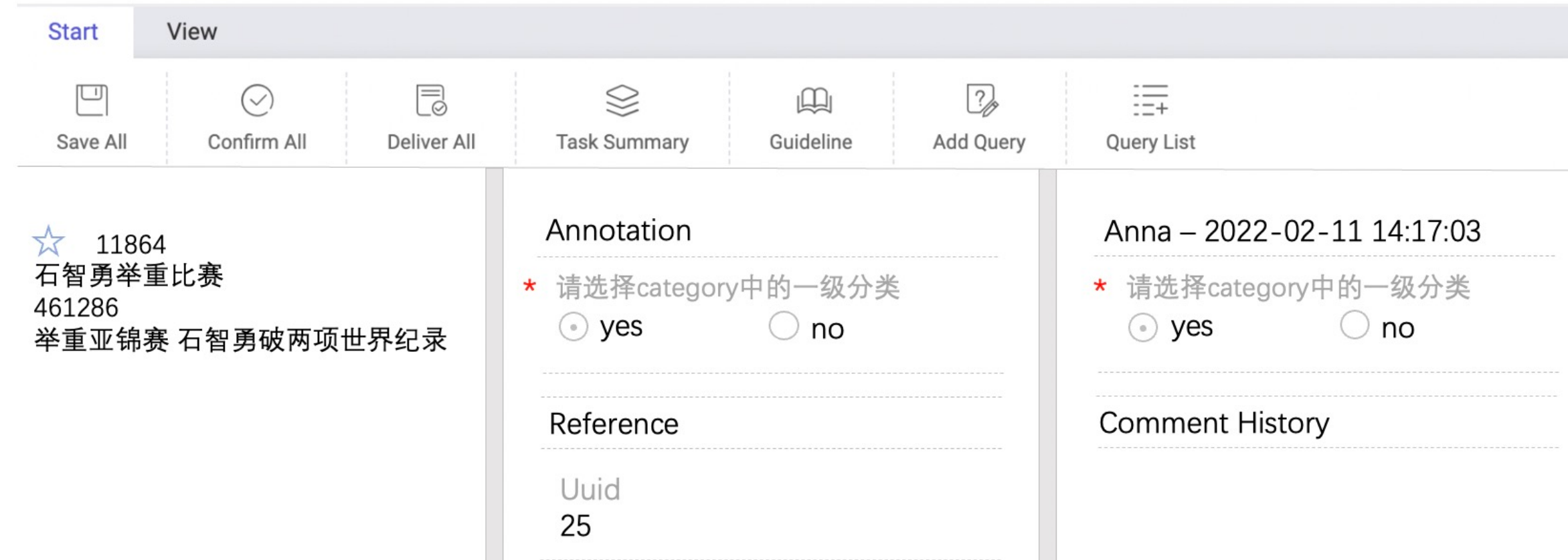}
  \caption{Illustration of the annotation platform used. It mainly consists of three parts. Left: display the query-passage pair for annotating; Middle: annotators can choose the label result; Right: the expert can check whether the label result is correct, and add some comments if needed.}
\end{figure*}

\section{Data Construction}
\subsection{Data Collection}
The core of constructing a passage retrieval dataset is to build high quality query-passage relevant pairs. To generate related query-passage pairs, we first sample some queries from the search logs of different search systems in Alibaba group. We attempt to filter out potentially relevant query-passage pairs based on user behaviors. It should be noted that not all passages clicked by the user are semantic relevant to a search query. For a commercial search engine system, the results finally displayed to users are not only decided by semantic relevance but also depended on excess features such as personalization and item popularity. For Multi-CPR, we only consider the semantic relevance between the query and passage. Therefore, to ensure the quality of the final dataset, we filter out the query-passage pairs with a relatively low number of clicks. Finally, human annotators will annotate all selected pairs to determine whether each pair is semantically related. 

\begin{table}[t]
\caption{An example of annotated sample with multiple passages for one query in E-commerce domain.}
\label{multi-passage-sample}
\renewcommand\arraystretch{1.0}
\begin{tabular}{l|p{5.3cm}|c}
\toprule
\textbf{Query} & 阔腿裤女冬牛仔 (\color[HTML]{3166FF}{Women's winter wide leg pants}) & Label \\
\midrule
\textbf{Passage-1} & 阔腿牛仔裤女秋冬款潮流百搭宽松 ( \color[HTML]{3166FF}{Women's wide leg jeans for the autumn/winter season, stylish and easy to match}) & Yes \\
\midrule
\textbf{Passage-2} & 牛仔阔腿裤女大码胖mm高腰显瘦夏季薄款宽松垂感泫雅拖地裤子   (\color[HTML]{3166FF}{Women's wide leg jeans for the autumn/winter season, stylish and easy to match, large size for fat mm, high-waisted and slim model look thin, suitable for summer,and same tall trousers with Xuanya}) & No \\
\midrule
\textbf{Passage-3} & 阔腿裤男大码高腰宽松 (\color[HTML]{3166FF}{Men's wide leg pants, large size, loose and high-waisted}) & No \\
\bottomrule
\end{tabular}
\end{table}


\begin{table*}[t]
\caption{Examples of query-passage pairs and annotated results of three different domains.}
\centering
\begin{tabular}{c|p{3.5cm}|p{4.5cm}|c|p{5cm}}
\toprule
& Query & Passage & Lable & Interpretation \\ \midrule
\multirow{2}{*}{E-commerce} & iphone13   & iphone13手机 (\color[HTML]{3166FF}{iphone13 mobile phone})  & Yes   & Product category word and product model word match \\ \cmidrule(l){2-5} & iphone13   & iphone13手机壳 (\color[HTML]{3166FF}{iphone13 mobile phone cases}) & No & Product category mismatch ( ``mobile phone'' vs ``mobile phone cases'' \\ \midrule
\multirow{2}{*}{\begin{tabular}[c]{@{}c@{}}Entertainment\\ video\end{tabular}} & 十六步交谊舞 (\color[HTML]{3166FF}{Sixteen-step social dance})  & 交谊舞四步 (\color[HTML]{3166FF}{Four-step social dance}) & No    & Attribute word mismatch (Sixteen-step vs Four-step) \\ \cmidrule(l){2-5}  & 十六步交谊舞 (\color[HTML]{3166FF}{16 steps ballroom dance})  & 广场舞16步对跳 (\color[HTML]{3166FF}{Sixteen-step Square dance by pair})  & Yes   & Attribute match and Dance type match \\ \midrule
\multirow{2}{*}{Medical}  & 宝宝腹胀的原因是什么 (\color[HTML]{3166FF}{What are the causes of abdominal distension for babies?}) & 脑梗塞是由于脑部的缺血缺氧 引起的脑组织的坏死及软化，常见的有脑血栓及脑栓塞 (\color[HTML]{3166FF}{Cerebral ischemic stroke is the necrosis and softening of brain tissue caused by ischemia and hypoxia in the brain, commonly known as cerebral thrombosis and cerebral embolism.}) & No & symptom word mismatch (``abdominal bloating'' vs ``cerebral ischemic stroke'' \\ \cmidrule(l){2-5}  & 宝宝腹胀的原因是什么 (\color[HTML]{3166FF}{What are the causes of abdominal distension for babies?}) & 宝宝出现腹胀都是因为消化不太好或是着凉的原因引起的 (\color[HTML]{3166FF}{The baby's abdominal bloating is caused by poor digestion or cold.})  & Yes & symptom word match  \\ \bottomrule
\end{tabular}
\end{table*}

\subsection{Data Annotation}
\label{section:data-annotation}
As mentioned above, the Multi-CPR dataset covers three different specific domains. Naturally, queries in each domain are sampled from different search systems. In specific, queries in the domain of E-commerce, Entertainment video and Medical are sampled from Taobao search\footnote{https://www.taobao.com}, Youku search\footnote{https://www.youku.com}, Quark search\footnote{https://www.myquark.cn} systems respectively. During the construction of the dataset, we seek to ensure the quality and practicability of the final produced dataset in the following aspects:

\subsubsection{Query Distribution}
For each domain, we sample queries from the search logs within a single day, and the selected queries are uniformly sampled based on all distinct queries. Such a sampling strategy avoids the sampled data being concentrated in high-frequency queries. Thus, the resulted passage retrieval models should also take into account the performance of long-tail queries. 

\subsubsection{Annotation Guideline} For each query-passage pair, our annotation process contains only one component, $i.e,$, to determine whether the query and the passage are truly semantically related. Since our query-passage pairs are sampled from search logs, there may be multiple relevant candidate passages for some queries, as illustrated in Table 2. For this kind of sample, we require the human annotators to compare all candidates and then mark the most semantically relevant passage as the positive, and the others as negatives. If there is no relevant passage in all candidates, then all candidates will be marked as negatives. For all domains, We summarize several basic principles to determine the relevance between the query and passage:

\vspace{0.1cm}
\noindent\textbf{Explicitness} For each candidate pair, we require that both query and passage are semantically completed. Moreover, the search intent of query is explicit. Taking the query-passage pair <query:\begin{CJK}{UTF8}{gbsn}电影(movie)\end{CJK}, passage: \begin{CJK}{UTF8}{gbsn}哈利波特与魔石(Harry Potter and the Magic Stone)\end{CJK}> as an example, the search concept of ``电影 (movie)'' is quite broad, and there are many docs that can meet this search requirement. On the contrary, for pair <\begin{CJK}{UTF8}{gbsn}哈利波特电影(harry potter movies)\end{CJK}, \begin{CJK}{UTF8}{gbsn}哈利波特与魔石(Harry Potter and the Magic Stone)\end{CJK}>, the query search intent is more explicit, and the doc is semantically complete. During the annotation process, query-passage pair that violates the explicitness principle will be eliminated directly.

\vspace{0.1cm}
\noindent\textbf{Headword Relevance} Headwords and central topic of query and passage should be closed. For example, for the pair <\begin{CJK}{UTF8}{gbsn}冬季阔腿裤女(women's winter wide leg pants), 冬季连衣裙女(women's winter dress)\end{CJK}. The headwords of query and passage are ``阔腿裤 (wide-leg pants)'' and ``连衣裙 (dress)'' respectively, which are totally different. For query \begin{CJK}{UTF8}{gbsn}冬季阔腿裤女 (women's winter wide leg pants)\end{CJK}, the passage \begin{CJK}{UTF8}{gbsn}阔腿裤牛仔冬季韩式女 (women's winter korean style wide leg denim pants)\end{CJK} is semantically related, since both shares the same headword.

\vspace{0.1cm}
\noindent\textbf{Full Matching} Passage contains a full answer to query rather than a partial answer. For query \begin{CJK}{UTF8}{gbsn}``鼻子上有黑头该怎么去除'' (How to get rid of blackheads on your nose)\end{CJK} in the medical domain, some passages only introduce the arsing of blackheads, but do not fully introduce how to get rid of blackheads. This type of passages will be marked as irrelevant.

Apart from the three universal principles introduced above, we also have specially designed principles for each domain by considering that each domain has its characteristics. Especially for the headwords relevance principle, the core factors with concern for each domain are different. For example, in the domain of E-commerce, the headwords are usually brand and category words, but in the domain of Entertainment video, they are usually names of the actor, roles or styles of the movie. In Table 3, we show some examples for each domain to more clearly illustrate our annotation guideline.

\subsubsection{Quality Control}
To ensure that each annotator can produce high quality annotations, we set a pre-annotation step. We first let the annotators read our instructions thoroughly and ask them to annotate a certain number of test samples (from $100$ to $200$). The expert examiners on this task checked whether the annotation satisfies the pre-defined annotation guideline. The annotators that meet all the principles can continue to annotate. After finishing all the labeling task, the expert examiners will sample $20\%$ of the annotators' data and check them carefully. If the acceptability ratio is lower than $95\%$, the corresponding annotators are asked to revise their annotations. The loop stops by the end when the acceptability ratio reaches $95\%$. We have an internal annotation platform (as illustrated in Figure $1$) to assist annotators and experts in producing datasets.

For some samples, we found that determining whether the query-passage pair is relevant is relatively subjective. It is possible that different annotators generate different labels for the same sample. We control the consistency of annotated labels by employing an inter-annotator agreement labeling methods. Concretely, for each sample, we gather at least $5$ independent annotators' annotation results, and the samples whose annotation results are more than $80\%$ agreement will be retained. 

\subsubsection{Passage Set Selection} Since our query-passage pairs are sampled from real search systems, it is impossible for us to release all passages in the search engine as the passage set given that the collection of passages is too large (on billion-level). Therefore, we attempt to build a passage set on a smaller scale, which is mainly consisting of two parts. Passages in query-passage pairs labeled as positive samples are bound to remain in the final passage set. At the same time, we will also uniformly sample passages from all passage collections to supplement the final set until the size of the final passage set reaches the number we expected. For three different domains, the number of the final passage set size is set at around 1 million. Such a uniform sampling strategy and dataset scale ensure the diversity of sampling passages and the passage retrieval efficiency simultaneously.

\subsection{General Domain Dataset}
In addition to the three domain datasets introduced above, we also construct a general domain passage retrieval dataset based on the existing open domain Chinese machine reading comprehension (MRC) dataset DuReader~\cite{he-etal-2018-dureader}. DuReader collects documents from the search results of Baidu Search\footnote{https://www.baidu.com/}. DuReader contains $200$K questions, $1$M documents and more than $420$K human-summarized answers, which is the largest Chinese MRC dataset so far. For each question, the Dureader dataset provides multiple documents which may contain the answer to the question. Each document consists of a title and body text. For each question and its associated documents, the original DuReader dataset has divided each document into independent short passages, and has marked whether each passage contains the correct answer or is semantically related to the question. 

To convert the MRC data format into a data format usable by the passage retrieval task, referring to the method described in MS MARCO~\cite{Campos2016MSMA}, for each question (query) in DuReader, we select passages containing the correct answer to building positive query-passage pair and take the union of all the passages in DuReader as the final passage set. 

\begin{table*}[t]
  \caption{Dataset statistics of three different domains.}
  \label{tab:daa-statistics}
  \begin{tabular}{ccccccc}
    \toprule
    Domain & Train & Test & Passages & Avg Length of Query & Avg Length of Passage \\
    \midrule
    General & 245897 & - & - & 9.56 & 85.38 \\
    E-commerce & 100000 & 1000 & 1002822 & 6.90 & 32.96 \\
    Entertainment video & 100000 & 1000 & 1000000 & 7.41 & 27.45 \\
    Medical & 99999 & 1000 & 959526	& 17.07 & 122.02 \\
    \bottomrule
\end{tabular}
\end{table*}

\subsection{Dataset Statistics}
Following previous work~\cite{Campos2016MSMA}, we only keep the positive query-passage samples in the final training and testing set. The overall statistics of the Multi-CPR dataset and the converted general domain dataset are summarized in Table \ref{tab:daa-statistics}. 


\section{Task and Experiments}
\subsection{Task Definition}
Given a query $q$, a passage retrieval model aims to recall all potentially relevant passages from a large corpus $\mathcal{C}=\{p_1, p_2,...,p_N\}$. Thus, the passage retrieval task can be formulated as:
\begin{equation}
    \mathcal{R}:(q,\mathcal{C}) \rightarrow \mathcal{C}_r,
\end{equation}
which takes $q$ and $\mathcal{C}$ as input and then returns a much smaller set of passages $\mathcal{C}_r \subset \mathcal{C}$, where $|\mathcal{C}_r| \ll N$.

For the passage retrieval task, the most fundamental problem is to estimate the degree of relevance between a query $q$ and a passage $p$. Existing retrieval systems can be divided into two typical groups:

\noindent \textbf{Sparse Retrieval Medels} The key idea of these models is to utilize exact matching signals to design a relevance scoring function. Specifically, these models consider easily computed statistics (e.g., term frequency, inverse document frequency) of terms-matched signals between $q$ and $p$. And the relevance score is derived from the sum of contributions from each query term that appears in the passage. Among these models, BM25~\cite{robertson2009probabilistic} is the most representative and still be regarded as a strong baseline for passage retrieval task. 

\noindent \textbf{Dense Retrieval Medels} The key idea of these models is to leverage deep neural networks to convert text into continuous vector representation for relevance estimation. These models use the low-dimension representations of $q$ and $p$ as the input and are usually trained with scale annotated relevance labels. Compared with traditional sparse retrieval methods, these models can be trained without handcrafted features in an  end-to-end manner. Recently, due to the great success achieved by BERT in the natural language process field, many works adopt the BERT model as the encoder for query and passage to obtain the final representation so as to better compute the final relevance score.

\begin{figure*}[ht]
  \centering
  \label{fig:model-architecture}
  \includegraphics[width=\linewidth]{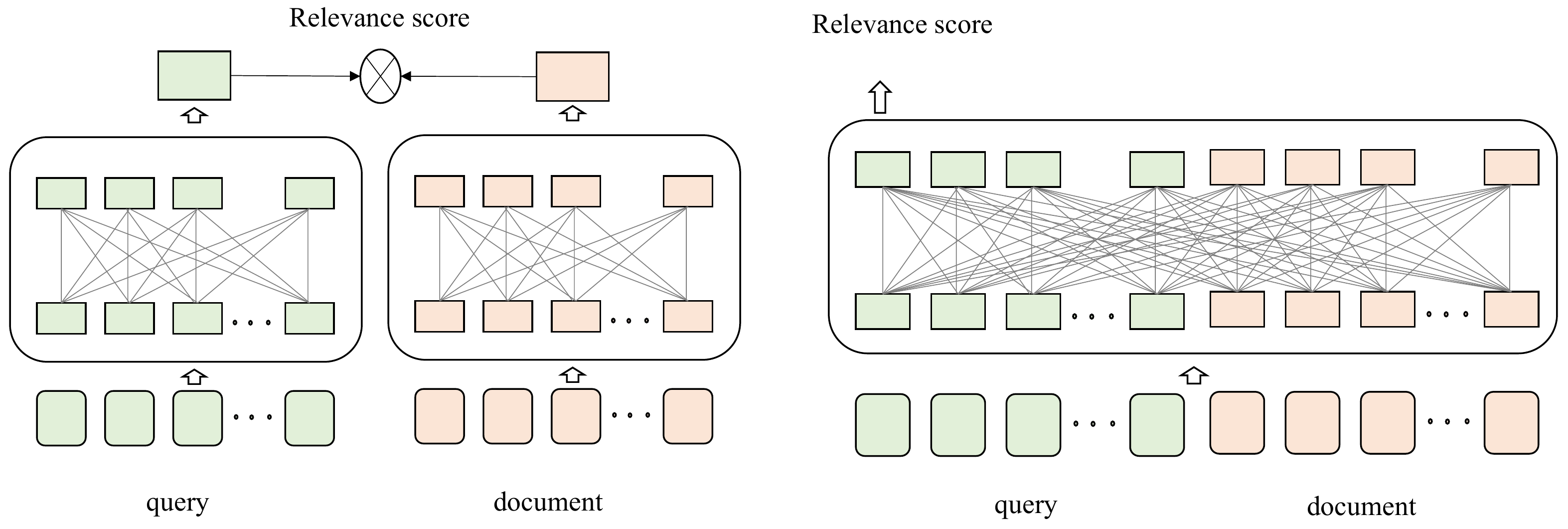}
  \caption{Illustration of BERT based passage retrieval and reranking model. Retrieval (left): query and passage are encoded independently by a dual-encoder. Reranking (right): query and passage are concatenated, jointly encoded by a cross-encoder.}
\end{figure*}

\subsection{Methods}
In this section, we will introduce some widely-adopted passage retrieval models (including both sparse and dense models) for experiments. 

\noindent{\textbf{BM25}} BM25 is the most widely used term-based passage
retrieval method. Practically, BM25 ranks a set of passages based on the query terms appearing in each passage, regardless of their proximity within the passage. 

\noindent{\textbf{Doc2Query}}~\cite{nogueira2019document} Doc2Query is still a term-based passage retrieval method. Doc2Query alleviates the term mismatch problem in the BM25 via training a neural sequence-to-sequence model to generate potential queries from passages, and indexes the queries as passage expansion terms. Different from the BM25 method, the implementation of the doc2query method relies on labeled query-passage pairs.

\noindent{\textbf{DPR}}~\cite{karpukhin2020dense} DPR is the most widely used dense passage retrieval method, which provides a strong baseline performance. It learns dense embeddings for the query and passage with a BERT-based encoder separately. The embeddings of query and passage are then fed into a ``similarity'' function to compute the final relevance score.

The retrieval performance of the DPR model is mainly determined by two factors: the BERT backbone network and the labeled query-passage dataset adopted. Therefore, in order to gain a deep understanding of the DPR model in domain passage retrieval, we conduct various settings based on the DPR model architecture by replacing the original BERT model with a BERT model that has continuously trained on in-domain raw text (DPR-2) or leveraging different domain labeled datasets to carry out the training process (DPR-1).  

\subsection{Evaluation Metrics}
Following the evaluation methodology used in previous work~\cite{Campos2016MSMA}, the retrieval performance is evaluated by Mean Reciprocal Rank at 10 passages (MRR@10) and recall precision at depth $1000$ (Recall@1k). For the passage reranking results, we only report the result of the MRR@10 metric.

\begin{table*}[t]
\caption{Results on three domain datasets. ``In-Domain'' indicates that the training dataset adopted is from the corresponding domain. ``BERT-CT'' notes that the BERT model is continuing trained with domain corpus.}
\label{tab:main-exp-results}
\resizebox{1.9\columnwidth}{!}{
\begin{tabular}{@{}c|c|c|c|cc|cc|cc}
\toprule
\multicolumn{2}{c|}{\multirow{2}{*}{Models}} &
  {\multirow{2}{*}{Dataset}} &
  {\multirow{2}{*}{Encoder}} &
  \multicolumn{2}{c|}{E-commerce} &
  \multicolumn{2}{c|}{Entertainment video} &
  \multicolumn{2}{c}{Medical} \\ \cmidrule(l){5-10} 
\multicolumn{2}{c|}{} & & &MRR@10 & Recall@1000 & MRR@10 & Recall@1000 & MRR@10 & Recall@1000 \\ \midrule
\multirow{2}{*}{Sparse} & BM25 & - & - & 0.2253 & 0.8150 & 0.2252 & 0.7800 & 0.1869 & 0.4820 \\
 & Doc2Query & - & - & 0.2385 & 0.8260 & 0.2378 & 0.7940 & 0.2095 & 0.5050 \\ \midrule
\multirow{3}{*}{Dense} & 
 DPR & General & BERT & 0.2106 & 0.7750 & 0.1950 & 0.7710 & 0.2133 & 0.5220 \\
 & DPR-1 & In-Domain & BERT & 0.2704 & 0.9210 & 0.2537 & 0.9340 & 0.3270 & 0.7470 \\
 & DPR-2 & In-Domain & BERT-CT & 0.2894 & 0.9260 & 0.2627 & 0.9350 & 0.3388	  & 0.7690 \\
 \midrule
\end{tabular}
}
\end{table*}

\subsection{Implementation Details}
For sparse retrieval methods, we adopt the pyseirni~\cite{lin2021pyserini} tool for experiments. For dense retrieval methods, we mainly focus on the DPR architecture. Following previous work, we use the Chinese BERT-base model released by Google Research\footnote{https://github.com/google-research/bert}. During the model training process, the in-batch negative optimization method is adopted with an initial learning rate of $1e-5$ and a batch size of 32. The Adam Optimizer~\cite{kingma2014adam} is adopted during the training process. All DPR models are trained on a single NVIDIA-V100 GPU. We use the faiss\footnote{https://github.com/facebookresearch/faiss} package for embedding indexing and searching.  

\subsection{Results}
The overall experimental results on the test set are shown in Table \ref{tab:main-exp-results}, from which we can observe that:

(1) On three different domain datasets in Multi-CPR, the retrieval performance of dense models outperforms sparse models. Taking the BM25 model and the DPR-1 model as an example, the average MRR@10 value over the three datasets are $0.2124$ and $0.2837$ respectively. The retrieval performance on the MRR@10 metric is largely improved by $33.57\%$, which points out the value of high-quality labeled data for the optimization of dense passage retrieval models.

\begin{table}[t]
\caption{Full ranking results of BERT reranking model on three domain datasets.}
\label{tab:res-full-ranking}
\resizebox{1.0\columnwidth}{!}{
\begin{tabular}{c|c|c|c|c}
\toprule
\multirow{2}{*}{Retrieval} & \multirow{2}{*}{Reranker} & E-commerce & \begin{tabular}[c]{@{}c@{}}Entertainment \\ video\end{tabular} & Medical \\ \cmidrule{3-5} 
      &      & MRR@10 & MRR@10 & MRR@10 \\ \midrule
BM25  & -    & 0.2253 & 0.2252 & 0.1869 \\ 
BM25  & BERT & 0.2784 & 0.3212 & 0.2673 \\ \midrule
DPR-1 & -    & 0.2704 & 0.2537 & 0.3270 \\ 
DPR-1 & BERT & 0.3624 & 0.3772 & 0.3855 \\ \bottomrule
\end{tabular}
}
\end{table}

(2) For sparse methods, the BM25 method provides a strong baseline for all the three domain datasets. Especially on the E-commerce dataset, the retrieval performance of BM25 is even slightly better than the DPR model (MRR@10: $0.2253$ vs $0.2106$, Recall@1000: 0.8150 vs 0.7750). We infer that the reason for this phenomenon is that the average length of query and passage in the domain of E-commerce is relatively short, and the search intent is explicit to some content. The method based on exact term matching can provide satisfactory retrieval results. This observation illustrates that traditional unsupervised term-based retrieval methods such as BM25 can still provide valuable results for passage retrieval in some specific domains. Moreover, as an optimization method that has been verified in previous work, Doc2Query has also achieved significant improvement on all three datasets as expected.

(3) For dense methods, we conduct the analysis from two aspects. In the dataset aspect, we can find that the DPR model trained on in-domain labeled dataset has achieved remarkable performance improvement compared to the dense model trained with general domain data, even though the size of labeled dataset is much smaller. As such, we can conclude that labeled data in general domain is helpful for training dense retrieval model on specific domain to some extent, but the in-domain labeled data could provide more effective and valuable information for model training. In the encoder aspect, by replacing the BERT model with a BERT model that has been continuing trained on in-domain raw text, the performance of DPR-2 model is significantly better than that of DPR-1. This phenomenon is in line with the observation in previous work~\cite{gururangan2020don}. Furthermore, we find that the method of BERT continuous training could achieve greater improvement in domains with larger discrepancies compared to the general domain (e.g, the medical domain).

(4) For both sparse and dense methods, we have attempted some existing optimization strategies based on the baseline model. It can be seen that the improvement brought by the optimization of the dense model is much larger than that of the sparse model. This once again shows that the dense model armed with labeled dataset has more space for optimization, which also shows the value of labeled data for domain specific Chinese passage retrieval task.

\section{Discussion and Analysis}
\label{section:Discussion}

\subsection{Full Ranking Performance}
Recent state-of-the-art passage retrieval systems are usually built with a multi-stage framework~\cite{nogueira2019passage,ren2021rocketqav2}, which consists of a first-stage retriever that efficiently produces a small set of candidate passages followed by one or more elaborate rerankers that rerank the most promising candidates. Similar to the dense retrieval methods, pre-trained language models have a major impact for reranking methods via providing rich deep contextualized matching signals between query and passage, as illustrated in Figure $2$. Here, in order to verify the practicability of the Multi-CPR dataset in both quality and scale, we also implement experiments with the BERT base reranking model (see Figure 2).

We first introduce some basic concepts of BERT base reranking model. We aim to train a BERT reranker to score each query passage pair:
\begin{equation}
    score(q,p) = W^{T}\textbf{cls}(BERT(concat(q,p))
\end{equation}
where cls extracts BERT's [CLS] vector and $W$ is a projection vector. Following previous work~\cite{Gao2021RethinkTO}, we optimize the reranking model with a contrastive learning objective. Concretely, for each query, we aggregate all negatives sampled from retrieval passage candidates. Thus, for each query $q$, we form a group $G_p$ with a single relevant positive $p^{+}$ and multiple negatives. By taking the scoring function defined in equation (2), the contrastive loss for each query $q$ can be denoted as:
\begin{equation}
    \mathcal{L}_q := -log\frac{exp(score(q, p^{+})}{\sum_{p\in G_p} exp(score(q, p))} 
\end{equation}

In Table \ref{tab:res-full-ranking}, we summarize the full ranking experiments results on Multi-CPR. From which we can observe that: 1) Reranking model can indeed improve the final passage retrieval performance. In statistics, the retrieval-then-re-ranking pipeline gets an average improvement of $32.5\%$ on the three datasets. 2) Better initial retrieval results can produce better raranking results. Intuitively, better retrieval results provide the BERT reranker with more quality negative passages, which are fruitful for the optimization of the contrastive loss as denoted in Equation (3).

The full ranking experiment results on Multi-CPR are in line with previous studies which leverage other existing datasets in English filed~\cite{padigela2019investigating,Gao2021RethinkTO}. These experiment results again prove that the Multi-CPR is qualified to build a passage reranking model for specific domain.


\begin{table*}[t]
\caption{Examples of top retrieval results of Sparse and Dense retrieval models.}
\label{tab:case-table}
\begin{tabular}{c|c|p{5.0cm}|p{5.0cm}}
\toprule
Domain &
  Query &
  Sparse Retrieval &
  Dense Retrieval \\ \midrule
\multirow{3}{*}{E-commerce} &
  \multirow{3}{*}{lining无界 (\color[HTML]{3166FF}{Lining boundless})} &
  李宁19春季款无界X情侣缓震训练鞋 (\color[HTML]{3166FF}{Li Ning 19 spring models boundless X couple cushioning training shoes}) &
  李宁无界缓震训练鞋2019夏秋款 (\color[HTML]{3166FF}{Li Ning boundless cushioning trainers 2019 summer and autumn models.}) \\ \cmidrule(l){3-4} 
 &
   &
  适配lining李宁 驭帅11 10... (\color[HTML]{3166FF}{Adapt to lining Li Ning Yu Shuai 11 10...}) &
  李宁春秋限量版男女训练减震一体织鞋套袜子运动鞋\begin{CJK}{UTF8}{bsmi}無\end{CJK}界 (\color[HTML]{3166FF}{Li Ning spring and autumn limited edition men and women training shock-absorbing one piece woven socks sneakers, boundless}) \\ \cmidrule(l){3-4} 
 &
   &
  lining李宁运动恢复颈椎按摩器 (\color[HTML]{3166FF}{lining Li Ning sports recovery cervical spine massager}) &
  李宁19春季款无界X情侣缓震训练(\color[HTML]{3166FF}{Li Ning 19 Spring Unbounded X Couples Cushioning Training}) \\ \midrule
\multirow{3}{*}{Medical} &
  \multirow{3}{*}{\begin{tabular}[c]{@{}c@{}}孩子嘴里擦紫药水\\ 产生副作用了怎么办呢\\ (\color[HTML]{3166FF}{What should be done if the} \\ \color[HTML]{3166FF}{kid has side effects from rubbing}\\ \color[HTML]{3166FF}{gentian violet in his mouth?})\end{tabular}}  &
  烫伤后正规治疗不使用紫药水与红药水,...,会产生副作用 (\color[HTML]{3166FF}{Regular treatment after scald does not use gentian violet and gentian violet,..., there will be side effects}) &
 不要紫药水，有问题不好观察，现在不主张用。 停药就好了。意见建议:多用白水漱口 （\color[HTML]{3166FF}{Do not use gentian violet, there is a problem that it is not well observed and is not advocated now. Stop the medicine will be fine. Using white water rinsing mouth is suggested.}）\\ \cmidrule(l){3-4} 
 &
   &
  紫药水孕妇要慎重使用... (\color[HTML]{3166FF}{Gentian violets should be used with caution by pregnant women}) &
  你好，一般紫药水是不能轻易给宝宝用的情况的 (
\color[HTML]{3166FF}{Hi, gentian violet are not suggested to be used for babies}) \\ \cmidrule(l){3-4} 
 &
   &
  对于炎症较轻、病程短的症状,可用紫药水或酒精消毒 (\color[HTML]{3166FF}{For mild inflammation and short duration of symptoms, disinfection with gentian violet or alcohol can be used}) &
  有关系的，可能是药物过敏引起的，最好还是停止使用， 还要注意孩子的饮食，以清淡为主的 (\color[HTML]{3166FF}{There is a relationship, it may be caused by drug allergies, it is best to stop using, and also pay attention to the child's diet, which should be mainly light food.}) \\ \bottomrule
\end{tabular}
\end{table*}

\subsection{Case Study}
In practical, to evaluate the relevance of a passage for a given query, retrieval models usually start from two aspects: 1) Precise term overlapping and 2) semantic similarity across related concept~\cite{luan2021sparse}. Usually, the sparse models excel at the first problem, while the dense models can be better at the second. To gain a deep understanding of the characteristics of sparse and dense retrieval models, here, we sample some queries along with their top retrieval results, as shown in Table \ref{tab:case-table}. 

For the query ``孩子嘴里擦紫药水产生副作用了怎么办'' (What should be done if the kid has side effects from rubbing gentian violet in his mouth?) in the medical domain, we can find that: 1) The headword ``紫药水'' (gentian violet) appears in all passages retrieved by the sparse model, although these passages do not completely match the query. In the retrieval results of dense models, the top-1 passage is the annotated golden passage, while the third passage does not contain the headword, and this passage is only semantically related to the query to some degree. Therefore, the sparse and dense models have very distinct characteristics, and can make different contributions to the overall passage retrieval performance. Some previous studies attempt to hybrid the two models to engage the merits of both for better retrieval performance~\cite{kuzi2020leveraging}. Our analysis illustrates that similar problems also exist in the Chinese passage retrieval task. We hope that the release of the Multi-CPR dataset can help to conduct more in-depth research on this problem.

Further, we find that different domains place different emphasis on the sparse and dense models. Queries in the E-commerce and entertainment video domains are relatively short in general. Although the search intent of the query is explicit, the key information is missing for some queries. In this case, the dense model can be helpful in finding semantically relevant passages by generalizing to larger concepts. Before the dense model was widely used, previous studies use query reformulation or synonym expansion to supplement the missing information in the query to improve the performance of the sparse model. For example, the query ``iphone13'' will be reformulated to ``iphone13 手机 (iphone13 mobile phone)'' in the E-commerce domain. The above observations show that armed with labeled dataset, the workload of designing handcrafted features in sparse models can be greatly eliminated, and the total complexity of the retrieval system can also be reduced.

\subsection{Availability}
We will publish the following resources in an open Multi-CPR repository\footnote{https://github.com/Alibaba-NLP/Multi-CPR}:
\begin{itemize}
    \item \textbf{Annotations}: All human annotated query-passage related pairs of three domains along with passage corpus will be released to the public.
    \item \textbf{Retrieval results}: For further studies, we will release the baseline retrieval results. 
    \item \textbf{BERT Models}: We will also provide the continuing trained BERT model with in-domain raw text for future studies.
    \item \textbf{Baselines}: we will release the source code to reproduce the baseline results presented in this paper.
\end{itemize}

\subsection{Future Directions}
We propose the following potential research questions to indicate future research directions on utilizing this Chinese passage retrieval dataset:

\noindent\textbf{1. Cross-domain Chinese passage retrieval.}

Cross-domain problem has been studied in many information retrieval~\cite{akkalyoncu-yilmaz-etal-2019-cross,ma2021zero} or natural language processing tasks~\cite{he2017unified,lin2018neural,ding2020coupling}. Commonly, models trained on one domain do not generalize well to another domain. Based on our experimental results in Table \ref{tab:main-exp-results}, we can observe that cross-domain is indeed a challenge for the Chinese passage retrieval task. Specifically, for two DPR models trained on dataset of general domain and in-domain, the DPR model trained on in-domain dataset has a $38.49\%$ lead on MRR@10. Therefore, current retrieval systems built for the general domain do not have good transferability. 

Based on our Multi-CPR dataset, research in two directions can be carried out: 1) Cross-domain from general domain to specific domain. As introduced in Section \ref{sec:related-work}, Chinese passage retrieval for the general domain has been studied for a relatively long period, and annotated dataset in the general domain is also available. How to leverage models and resources in the general domain to improve the retrieval performance in a specific domain is a problem worthy of study. 2) Cross-domain between specific domains. Apart from the three domains covered by Multi-CPR, there are other specific domains that can not be enumerated. As such, cross-domain research between different domains is also worth exploring.

\noindent\textbf{2.How to further improve in-domain Chinese passage retrieval?}

In our experiments, we find that in-domain retrieval performance can be greatly improved by using some of the optimization strategies proposed in previous work. For example, By continuing training the BERT model on domain raw text, the MRR@10 metric has increased by $25.06\%$ in the medical domain. Previously, due to the lack of a public labeled dataset, more in-depth research on the Chinese passage retrieval task has not been carried out. 

In the English field, many supervised optimization strategies have been proposed for both sparse models and dense models. Recently, as the dense models have shown greater advantages, each module of the dense model has been studied in depth. For the backbone network, except for the method of continuing training the BERT model on the domain corpus, various pre-trained language model architectures specially designed have achieved significant performance improvements on the multiple benchmark datasets (e,g. condenser~\cite{gao2021condenser}, coCondenser~\cite{gao2021unsupervised}). For the fine-tuning pipeline, based on the Multi-CPR dataset, we can conditionally verify whether the previously proposed methods are effective on the Chinese dataset (e,g. ANCE~\cite{xiong2020approximate}). Moreover, there is a big gap between Chinese and English. Thus, it is potential to explore more effective optimization strategies according to the characteristics of the Chinese passage retrieval task.

\noindent\textbf{3. Can other tasks benefit from Multi-CPR?}

In practice, passage retrieval is usually regarded as an intermediate step for the entire system. For example, for a web search system or recommendation system, passage retrieval is only one module in the whole process, and there are many subdivided upstream and downstream modules (e,g. query processor, CTR model, Re-ranking model, etc.). Here, we use the query processor module for a more detailed explanation. There are usually two basic modules contained in the query processor module: Query reformulation (QR) and Query suggestion (QS)~\cite{jansen2009patterns,sordoni2015hierarchical}. Concretely, the QR module aims to modify a query to improve the quality of search results to satisfy the user's information need, and the QS module aims to provide a suggestion that may be a reformulated query to better represent a user's search intent. There have been multiple research works attempting to build query reformulation models based on annotated passage retrieval dataset thus finally improving the retrieval performance~\cite{jansen2009patterns,sordoni2015hierarchical,dehghani2017learning,arabzadeh2021matches}. We believe that Multi-CPR also provides a solid data resource for similar research in the Chinese passage retrieval field. 

\section{Conclusion}

In this paper, we present Multi-CPR, a Chinese passage retrieval dataset that covers three specific domains. All queries and passages are collected from practical search systems and we present a detailed description of the entire dataset construction process. We develop a deep analysis of Multi-CPR and the experiments results of various competitive baselines further prove the challenge of our dataset. We also discuss some valuable research problems based on the Multi-CPR dataset for future work. Finally, we will open-source all the annotated datasets and related baseline codes.

\section*{ACKNOWLEDGMENTS}
We thank all anonymous reviewers for their helpful suggestions. We also thank all the annotators for constructing this dataset. Special thanks to Shuyi Li and Qiankun Sun for their efforts as expert examiners in the annotation process.

\clearpage
\bibliographystyle{ACM-Reference-Format}
\bibliography{chineseprdata}
\end{CJK}
\end{document}